\documentclass[english,12pt,aps,prd,a4paper,preprintnumbers,floatfix,nofootinbib,showpacs,superscriptaddress, notitlepage]{revtex4-1}
 \pdfoutput=1
\usepackage[usenames,dvipsnames]{color}  
\usepackage{graphicx}
\usepackage{caption}
\usepackage{subcaption}
\usepackage{amsmath}
\usepackage{amssymb}
\usepackage[colorlinks=true,citecolor=darkred,urlcolor=darkred, pdfborder={0 0 0}]{hyperref}
\usepackage[normalem]{ulem}

\usepackage{soul}
\definecolor{darkred}{rgb}{0.6,0,0}

\definecolor{linkcolor}{rgb}{0,0,0.5}

\def\gsim{\raise0.3ex\hbox{$\;>$\kern-0.75em\raise-1.1ex\hbox{$\sim\;$}}}
\def\lsim{\raise0.3ex\hbox{$\;<$\kern-0.75em\raise-1.1ex\hbox{$\sim\;$}}}

\def\beqn#1{\begin{equation}\label{#1}}
\def\eeqn{\end{equation}}

\def\beqa#1{\begin{eqnarray}\label{#1}}
\def\eeqa{\end{eqnarray}}

\def\Z2{$\mathcal{Z_2}$}

\newcommand {\ignore}[1]{}

\def\321{$\mathrm{SU(3) \otimes SU(2) \otimes U(1)}$ }

\newcommand{\AddrAHEP}{%
  AHEP Group, Institut de F\'{i}sica Corpuscular --
  CSIC/Universitat de Val\`{e}ncia, Parc Cient\'ific de Paterna.\\
 C/ Catedr\'atico Jos\'e Beltr\'an, 2 E-46980 Paterna (Valencia) - SPAIN}

\begin{document}

\bibliographystyle{unsrt}   

\title{\boldmath \color{BrickRed} Status and prospects of `bi-large' leptonic mixing}

\author{Gui-Jun Ding}\email[Email Address: ]{dinggj@ustc.edu.cn}
\affiliation{Interdisciplinary Center for Theoretical Study and Department of Modern Physics, \\
University of Science and Technology of China, Hefei, Anhui 230026, China}
\author{Newton Nath}
\email[Email Address: ]{newton@ihep.ac.cn}
\affiliation{
Institute of High Energy Physics, Chinese Academy of Sciences, Beijing, 100049, China}
\affiliation{
School of Physical Sciences, University of Chinese Academy of Sciences, Beijing, 100049, China}
\author{Rahul Srivastava}
\email[Email Address: ]{rahulsri@ific.uv.es}
\affiliation{\AddrAHEP}
\author{Jos\'{e} W. F. Valle}
\email[Email Address: ]{valle@ific.uv.es}
\affiliation{\AddrAHEP}

\begin{abstract}

\vspace{1.0cm}
{\noindent

Bi-large patterns for the leptonic mixing matrix are confronted with current neutrino oscillation data. We analyse the status of these patterns and determine, through realistic simulations, the potential of the upcoming long-baseline experiment DUNE in testing bi-large \emph{ansatze} and discriminating amongst them.
}
\end{abstract}
\preprint{USTC-ICTS-19-07}
\maketitle


\section{Introduction}


The lepton mixing matrix is the leptonic analogue of the Cabibbo-Kobayashi-Maskawa (CKM) matrix. It is given as $ U = U^{\dagger}_l U_{\nu}$ where $U_l$ and $U_{\nu} $ represent the diagonalization matrices corresponding to the charged-leptons and neutrino sectors, respectively.
The peculiar form taken by its matrix elements, which follows from current experimental data, has puzzled theorists ever since oscillations were established.
Even more so after reactor experiments established that $\theta_{13}$ is nonzero \cite{An:2012eh,An:2016ses}, and T2K has provided a hint for a nearly maximal value of the neutrino oscillation CP phase $\delta $~\cite{Abe:2017uxa}.
Many recent attempts to account for non-zero $\theta_{13}$ have been proposed~\cite{Minakata:2004xt,Albright:2010ap,Chen:2015siy,Pasquini:2016kwk,Chen:2018lsv}, including deviations from the conventional tribimaximal neutrino mixing pattern~\cite{harrison:2002er,Rahat:2018sgs}, that can provide realistic descriptions of neutrino oscillation data~\cite{Chen:2018eou,Chen:2018zbq}.

It is well-known that the magnitudes of the elements of the quark mixing matrix can be nicely described in terms of the Cabibbo angle~\cite{Tanabashi:2018oca}.
Motivated by the fact that the smallest lepton mixing angle is similar in magnitude to the largest of the quark mixing angles, it has been
suggested that the Cabibbo angle may act as the universal seed for quark and lepton mixings. This idea prompted alternative approaches
for describing the structure of the lepton mixing matrix involving bi-large neutrino mixing~\cite{Boucenna:2012xb,Ding:2012wh,Roy:2014nua}.
Specially constraining variants of this approach have been recently suggested~\cite{Chen:2019egu}.
Bi-large neutrino mixing implies that the lepton and quark sectors may be related with each other, possibly serving as a new starting point in the quest for quark-lepton symmetry and unification.

Here we examine the status of bi-large proposals for the leptonic mixing matrix by confronting them with current neutrino oscillation data. They require, in addition to the bi-large assumption on the matrix $U_{\nu}$ representing the neutrino diagonalization matrix, also the charged-leptons correction factor $U_l$ which we take to be ``CKM like'', motivated from Grand Unified Theory (GUT).
By carefully simulating upcoming long baseline oscillation experiments we also determine their potential in testing the bi-large \emph{ansatze}.
For definiteness we focus on the Deep Underground
Neutrino Experiment (DUNE) \cite{Acciarri:2015uup,Alion:2016uaj}. In order to determine its potential to test various bi-large predictions we simulate the experimental features according to their design specifications. What we describe in this letter follows the same general strategy as other symmetry-based studies in the context of DUNE such as those in Refs.~\cite{Chatterjee:2017ilf,Srivastava:2017sno,Srivastava:2018ser,Chakraborty:2018dew,Nath:2018xkz, Nath:2018fvw}, but focusing on the study of bi-large lepton mixing.
We determine the sensitivity regions of oscillation parameters within different bi-large \emph{ansatze} for the lepton mixing matrix. We also discuss the possibility of discriminating amongst different bi-large options.


\section{Before and after DUNE: the general case}
\label{sec:before-after-dune}

The proposal and technical details of the next generation superbeam neutrino oscillation experiment DUNE are described in~\cite{Acciarri:2015uup, Alion:2016uaj}.
The collaboration plans on using the Long-Baseline Neutrino Facility (LBNF) beam, which will be constructed over the next few years at Fermilab as a neutrino source.
The first detector will record particle interactions near the beam source, at Fermilab, while the second, much larger, underground detector
at the Sanford Underground Research Facility (SURF) in South Dakota, at about 1300 km distance, will use four 10~kton volume of liquid argon time-projection chambers (LArTPC). The expected design flux corresponds to 1.07 MW beam power which gives $1.47\times 10^{21} $ protons on target
per year for an 80 GeV proton beam energy.
For our numerical DUNE simulation we use the \texttt{GLoBES} package \cite{Huber:2004ka, Huber:2007ji} along with the auxiliary files in Ref.~\cite{Alion:2016uaj}.
We assume 3.5 years running time in both neutrino and antineutrino modes with a 40 kton detector volume.

\begin{figure}[!htbp]
\centering
\begin{center}
%
\includegraphics[height=7cm,width=8cm]{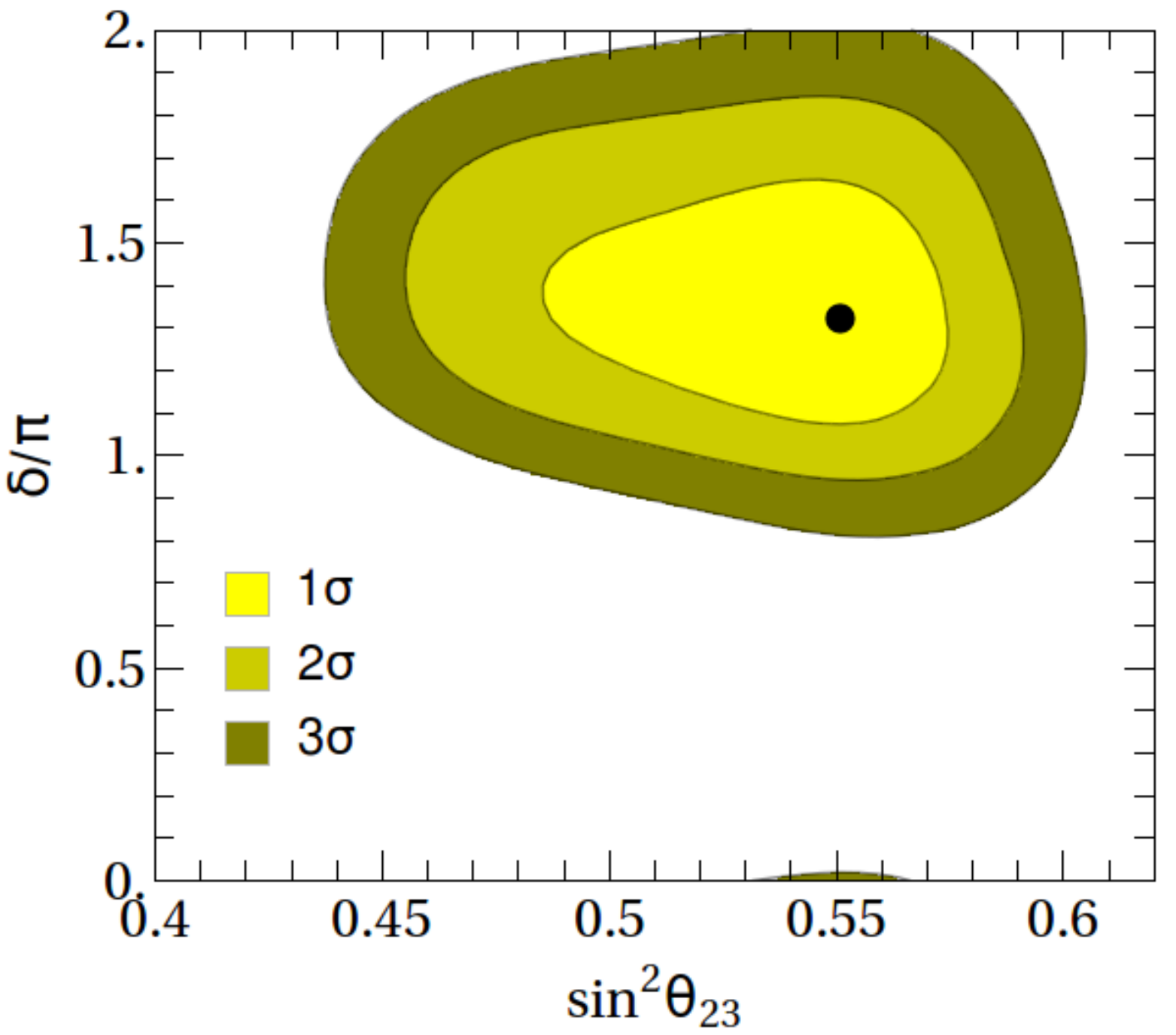}
\includegraphics[height=7cm,width=8cm]{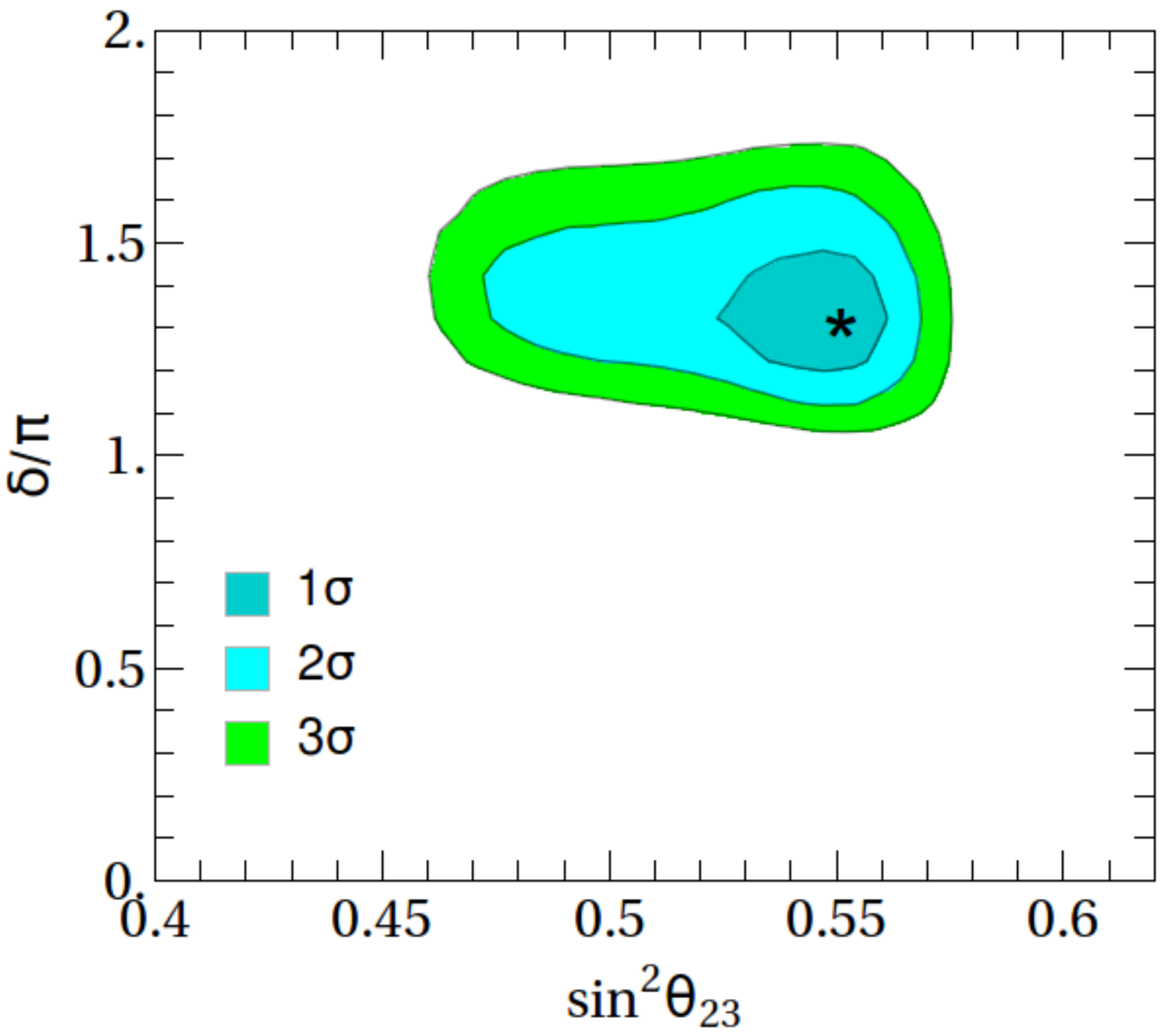}
 \end{center}
\caption{\footnotesize Current oscillation data (left panel): the filled regions indicate the allowed values in the $\sin^2\theta_{23} - \delta $ plane from the latest global-fit~\cite{deSalas:2017kay}, with the current best-fit point shown by a black dot.
The sensitivity expected in 3.5+3.5 DUNE running time is indicated in the right panel, assuming the current B.F.P. as the true value in the simulation.
}
\label{fig:gen}
\end{figure}

We also take into account both the appearance and disappearance channels of neutrinos and antineutrinos in the  numerical simulation. Both the signal and background normalization uncertainties for the appearance as well as disappearance channels  have been adopted in our analysis as mentioned in the DUNE CDR \cite{Alion:2016uaj}.
Furthermore, as the latest global analysis of neutrino oscillation data tends to favor normal mass hierarchy  (i.e., $\Delta m^2_{31} > 0$) over inverted  mass hierarchy (i.e., $\Delta m^2_{31} < 0$) at more than 3$ \sigma $ \cite{deSalas:2017kay, deSalas:2018bym}, we focus on the first scenario throughout this work.
The remaining numerical details that have been adopted here are same as \cite{Nath:2018fvw} unless otherwise mentioned.
Before considering the bi-large schemes we first recall the potential of the DUNE setup in probing the neutrino oscillation parameters within a generic oscillation
scenario. The results of our simulation are summarized in Fig.~\ref{fig:gen}.

\section{Bi-large Mixing Patterns}
\label{sec:BiLar}

As already mentioned, in contrast to quark mixing, the observed pattern of lepton mixing is described by two large mixing angles characterizing solar and atmospheric oscillations.
A specially intriguing possibility is provided by bi-large mixing \textit{ansatze}~\cite{Boucenna:2012xb}.
Apart from fitting observation, another reason for these \textit{ansatze} is that they fit well within the framework of unified theories~\cite{Ding:2012wh}. In addition, they open the door to the possibility of having the Cabibbo angle as a universal seed for quark as well as lepton mixings~\cite{Roy:2014nua}.
Here we give an overview of the predictions of bi-large mixing patterns for the lepton mixing matrix. For definiteness we focus on pattern {\bf T1}, {\bf T2} from the recent paper in~\cite{Chen:2019egu}, while {\bf T3} and {\bf T4} are taken from a previous paper~\cite{Roy:2014nua}. Following Ref.~\cite{Roy:2014nua} we assume that the bi-large patterns arise from the simplest Grand Unified Theories, where the charged-lepton and the down-type quarks have roughly the same mass.

\subsection*{Type-1 (T1)}


The first bi-large case {\bf T1} assumes that the neutrino mixing angles are related with the Cabibbo angle
as follows~\cite{Chen:2019egu}
\begin{equation}
\sin\theta_{23} = 1 - \lambda, ~~\sin\theta_{12} = 2\lambda, ~~\sin\theta_{13} = \lambda \;.
  \end{equation}
Truncated to $\mathcal{O} (\lambda^2)$, the neutrino part of mixing matrix $U_{BL1}$ is given by
\begin{align}
U_{BL1}\approx \left[
\begin{array}{ccc}
 1-\frac{5 \lambda ^2 }{2} & 2 \lambda & - \lambda  \\
\lambda - 2\sqrt{2} \lambda^{3/2}& \sqrt{2 \lambda} - \dfrac{\lambda^{3/2}}{2\sqrt{2}} & 1 - \lambda - \dfrac{\lambda^2}{2} \\
2 \lambda + \sqrt{2} \lambda^{3/2} & -1 + \lambda & \sqrt{2 \lambda} - \dfrac{\lambda^{3/2}}{2\sqrt{2}} \\
\end{array}
\right]\;.
\end{align}
In order to construct $ U_{BL1} $ we assume that the phase $ \delta = \pi $.
Notice that the observed leptonic mixing angles are obtained once the charged-lepton corrections have been incorporated (see discussion below).
Clearly $U_{BL1}$ on its own cannot be the full leptonic mixing matrix. In order to get the leptonic mixing matrix consistent with current oscillation data we need to take into account appropriate charged-lepton corrections to $U_{BL1}$. The $SO(10)$ GUT-motivated, CKM-type
charged-lepton corrections for this case are given by
\begin{align}\label{eq:Ul1}
U_{l_1}&=R_{23}(\,\theta_{23}^{CKM}\,)\Phi R_{12}(\,\theta_{12}^{CKM}\,)\Phi^{\dagger}\simeq \begin{bmatrix}
1-\frac{1}{2}\lambda^2 & \lambda\, e^{-i \phi} & 0 \\
-\lambda\, e^{i \phi} & 1-\frac{1}{2}\lambda^2 & A\lambda^2 \\
A \lambda^3 e^{i \phi} & -A \lambda^2 & 1
\end{bmatrix}.
\end{align}
Here, $\sin\theta^{\rm CKM}_{12}  =\lambda$ and $ \sin\theta^{\rm CKM}_{23}=A\lambda^{2}$, where
$\lambda=0.22453\pm0.00044$, $A=0.836\pm0.015$ are the Wolfenstein parameters~\cite{Tanabashi:2018oca},
and $\Phi = {\rm diag} \{ e^{-i\phi/2}, e^{i\phi/2},1\}$ with $\phi$ being a free parameter.
The lepton mixing matrix for \textbf{T1} is simply $U = U^{\dagger}_{l_1}  U_{BL1}$, with the mixing angles and the Jarlskog invariant $J_{CP} $, given by
\begin{equation}
\label{eq:BL1}
\begin{aligned}
\sin^{2}\theta_{13} & \simeq 4 \lambda^{2} (1 - \lambda) \cos^2\dfrac{\phi}{2} \;, \\
\sin^{2}\theta_{12} & \simeq 2 \lambda^{2} (2 - 2\sqrt{2\lambda} \cos\phi + \lambda) \;, \\
\sin^{2}\theta_{23} & \simeq (1 - \lambda)^{2} - 2\sqrt{2} A\lambda^{5/2} - 2\lambda^{3}(1+2\cos\phi) \;, \\
J_{CP}  & \simeq -  2\left(\sqrt{2}+\sqrt{\lambda}\right)\lambda^{5/2} \sin\phi \,.
\end{aligned}
\end{equation}
Note the presence of the next-to-leading order term $-2\lambda^3\sin\phi$ in $J_{CP}$.

\subsection*{Type-2 (T2)}

The second bi-large case {\bf T2} assumes that the neutrino mixing angles are related with the Cabibbo angle as follows~\cite{Chen:2019egu}
\begin{equation}
\sin\theta_{23} = 3\lambda, ~~ \sin\theta_{12} = 2\lambda, ~~\sin\theta_{13} = \lambda \;.
\end{equation}
The neutrino diagonalization matrix is approximately of the form
\begin{align}
U_{BL2}\approx \left[
\begin{array}{ccc}
 1-\frac{5 \lambda ^2 }{2} & 2 \lambda & - \lambda  \\
-2\lambda + 3\lambda^{2} & 1 - \dfrac{13 \lambda^{2}}{2} & 3\lambda \\
\lambda + 6 \lambda^{2} & -3\lambda + 2\lambda^{2} & 1 - 5\lambda^{2}\\
\end{array}
\right] \;.
\end{align}
As in the previous $ U_{BL1} $ case, in order to construct $ U_{BL2} $ we assumed $ \delta = \pi $.
Here, motivated by the  $ SU(5) $ unification, we take the contribution from the charged-lepton diagonalization matrix to be of the form
\begin{align}\label{eq:Ul2}
U_{l_2}& = \Phi^{\dagger} R^{T}_{12}(\,\theta_{12}^{CKM}\,)\Phi R^{T}_{23}(\,\theta_{23}^{CKM}\,)\simeq \begin{bmatrix}
1-\frac{1}{2}\lambda^2 & -\lambda\, e^{i \phi} & A \lambda^3 e^{i \phi} \\
\lambda\, e^{- i \phi} & 1- \frac{1}{2}\lambda^2 & - A\lambda^2 \\
0 & A \lambda^2 & 1
\end{bmatrix}\,, \end{align}
where $\theta_{12}^{CKM}$, $\theta_{23}^{CKM}$ and $\Phi$ take the same form as defined after Eq.~\eqref{eq:Ul1}.
Given the forms of $U_{BL2}$ and $U_{l_2}$, one can construct the lepton flavor mixing matrix for \textbf{T2} as $U=U^{\dagger}_{l_2} U_{BL2}$. The mixing angles and the Jarlskog invariant $J_{CP}$ are given as
\begin{equation}
\label{eq:BL2}
\begin{aligned}
\sin^{2}\theta_{13} & \simeq  \lambda^{2} - 6 \lambda^3 \cos\phi  + 8\lambda^4 \,, \\
\sin^{2}\theta_{12} & \simeq  \lambda^{2} (5 + 4 \cos\phi )  - 2\lambda^4(8+13\cos\phi) \,, \\
\sin^{2}\theta_{23} & \simeq 9\lambda^{2} + 6\lambda^{3}(A
+ \cos\phi)  - \lambda^4(8-2A\cos\phi-A^2)\,, \\
J_{CP}  & \simeq -\left[3+(16+A)\lambda\right]\lambda^{3} \sin\phi \;.
\end{aligned}
\end{equation}
Here again we have included the next-to-leading order term $-(16+A)\lambda^4\sin\phi$ in $J_{CP}$, which is comparable with the leading order term.


\subsection*{Type-3 (T3)}


Both patterns {\bf T3} as well as {\bf T4} have a common assumption~\cite{Roy:2014nua}
\begin{equation}
\sin\theta_{12}=\sin\theta_{23} =\psi\lambda, \qquad \sin\theta_{13}=\lambda\,.
\end{equation}
where $\psi $ is a free parameter whose value can be fitted from neutrino oscillation data. The cases {\bf T3} and {\bf T4} differ from each other in the type of charged-lepton corrections used. The exact form of the neutrino diagonalization matrix $U_{BL3}$ is
\begin{align}
 U_{BL3}=\left[
\begin{array}{ccc}
c\sqrt{1-\lambda^2}   ~&~   \psi\lambda\sqrt{1-\lambda^2}  ~&~  \lambda \\
-c\psi\lambda(1+\lambda)  ~&~   c^2-\psi^2\lambda^3   ~&~  \psi\lambda\sqrt{1-\lambda^2}  \\
-c^2\lambda+\psi^2\lambda^2  ~&~  -c\psi\lambda(1+\lambda)  ~&~  c\sqrt{1-\lambda^2}
\end{array}
\right],\quad c\equiv\cos \sin^{-1}(\psi\lambda)\,.
\label{eq:Ubl3}
\end{align}
In order to determine the lepton mixing matrix for this case here we choose $ \delta = 0 $.
For the case {\bf T3} the charged-lepton diagonalization matrix is taken to be same as the $U_{l_1}$
matrix of Eq.~\eqref{eq:Ul1}.
The resulting expressions of the mixing parameters are
\begin{equation}
\label{eq:bl3}
\begin{aligned}
\sin^2\theta_{13}&\simeq\lambda ^2-2 \psi\lambda^3\cos\phi+\left(-1+2Ac\cos\phi+\psi^2\right)\lambda^4\,, \\
\sin^2\theta_{12}&\simeq\left(c^4-2c^2\psi\cos\phi+\psi^2\right)\lambda^2+\left(c^4-\psi^2\right)\lambda^4\,, \\
\sin^2\theta_{23}&\simeq\psi^2\lambda^2+2\left(\cos\phi-A c\right)\psi\lambda^3+\left(1-\psi^2-2Ac\cos\phi+A^2c^2\right)\lambda^4\,,\\
J_{CP}&\simeq c^2\left[\psi+\left(\psi-Ac-\psi^3\right)\lambda\right]\lambda^3\sin\phi \;.
\end{aligned}
\end{equation}


\subsection*{Type-4 (T4)}

The {\bf T4} case also assumes the neutrino part of the bi-large mixing matrix to be the same as $ U_{BL3} $, as given by Eq.~\eqref{eq:Ubl3}. The  {\bf T4} case differs from the {\bf T3} case in the type of correction assumed for the charged-lepton diagonalization matrix.
In this case, the latter has the form $U_{l_2}$ given in Eq.~\eqref{eq:Ul2}.
The corresponding expressions for the mixing parameters are
\begin{equation}
\label{eq:bl4}
\begin{aligned}
\sin^2\theta_{13}&\simeq\lambda^2+2 \psi\lambda^3\cos\phi-\left(1-\psi^2\right)\lambda^4\,, \\
\sin^2\theta_{12}&\simeq\left(c^4+2c^2\psi\cos\phi+\psi^2\right)\lambda^2+\left(c^4-\psi^2\right)\lambda^4\,, \\
\sin^2\theta_{23}&\simeq\psi^2\lambda^2+2\left(Ac-\cos\phi\right)\psi\lambda^3+\left(1-\psi^2-2Ac\cos\phi+A^2c^2\right)\lambda^4\,,\\
J_{CP}&\simeq c^2\left[\psi+\left(\psi+Ac-\psi^3\right)\lambda\right]\lambda^3\sin\phi \;.
\end{aligned}
\end{equation}

After introducing the various types of bi-large mixing patterns, we now proceed to study their current status in the next section.


\subsection*{Current Status of bi-large Mixing Schemes }\label{sec:current}


\begin{figure}[h!tbp]
\centering
\begin{subfigure}[b]{1\textwidth}
\includegraphics[height=8cm,width=10cm]{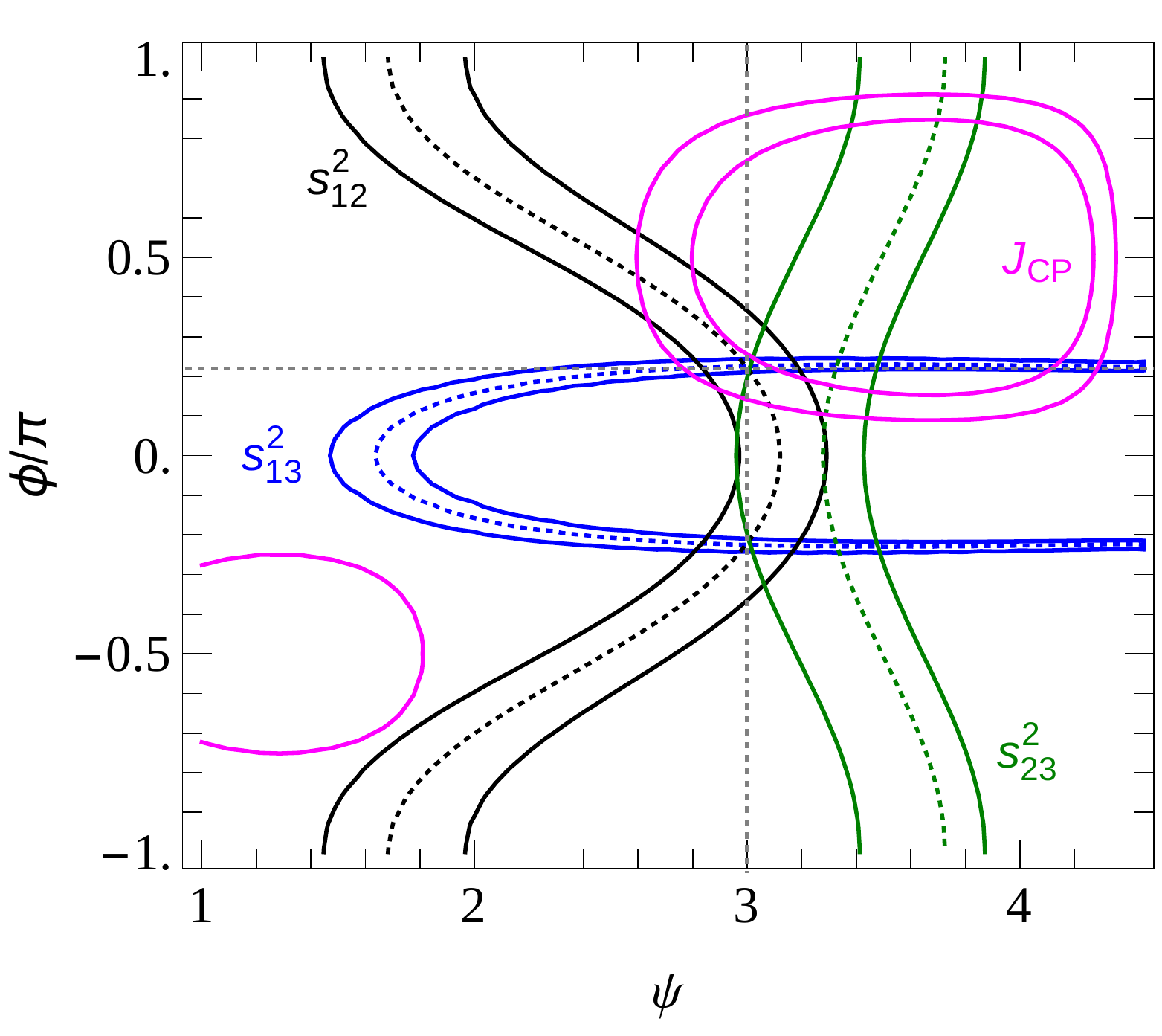}\captionsetup{justification=centering,margin=1cm}
\caption{\footnotesize The allowed parameter ranges of $\phi$ and $\psi$ for {\bf T3} ansatz.}
\end{subfigure}
\vspace{.005cm} \\
\begin{subfigure}[b]{1\textwidth}
\includegraphics[height=8cm,width=10cm]{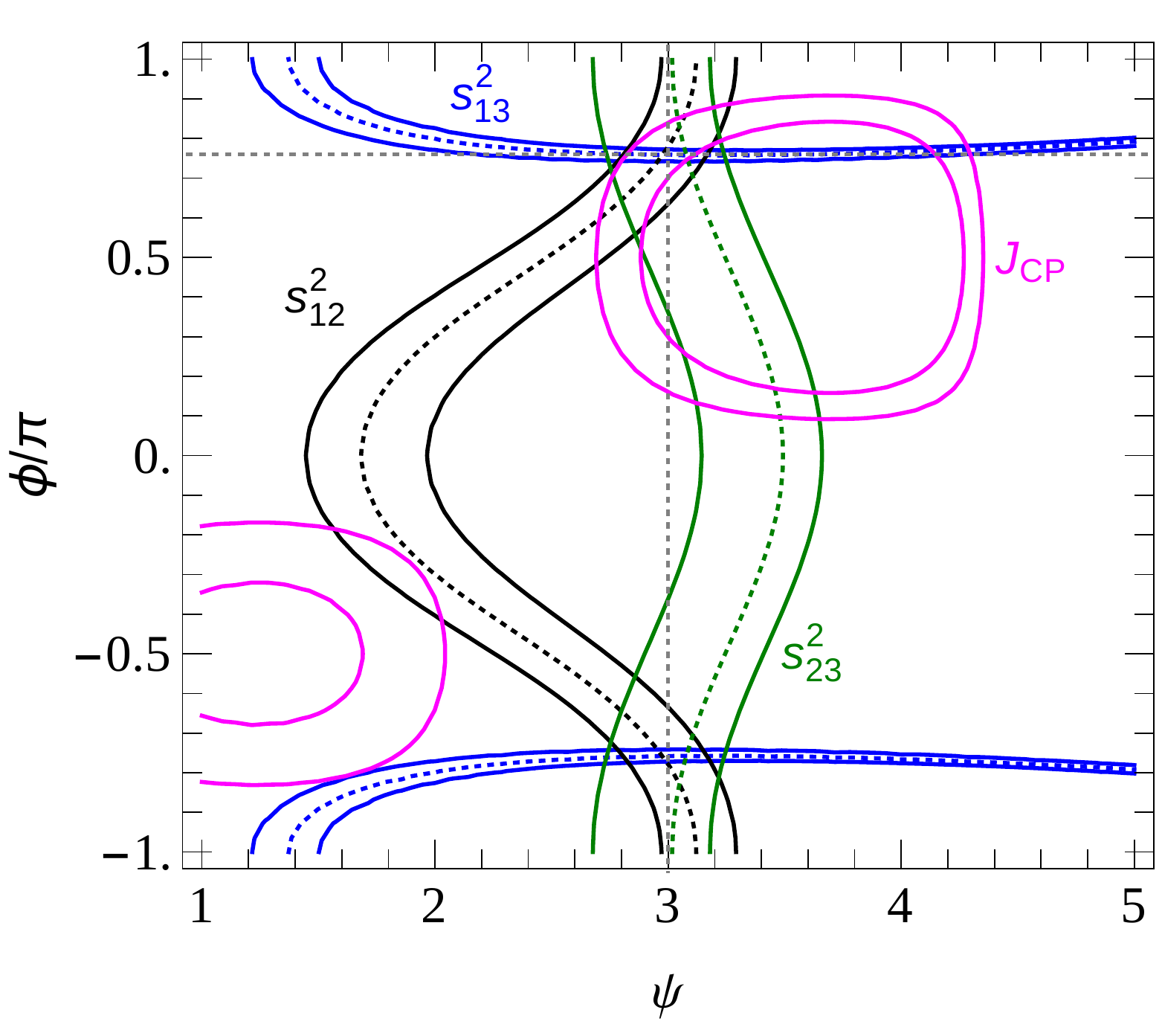}
\captionsetup{justification=centering,margin=1cm}
\caption{\footnotesize The allowed parameter ranges of $\phi$ and $\psi$ for {\bf T4} ansatz.}
\end{subfigure}
\caption{\footnotesize
Ranges of the mixing parameters are taken from the latest global neutrino oscillation
fit~\cite{deSalas:2017kay}. Here $s^2_{ij}$ denotes $\sin^2 \theta_{ij}$; $i,j = 1,2,3$ and $i \neq j$.
The dashed line in each $s^2_{ij}$  denotes the allowed values of
$\phi$ and $\psi$ corresponding to the current global best-fit points of $s^2_{ij}$.
The solid lines in each case denote the current 3$\sigma$ range.  }
\label{fig:phi-psi}
\end{figure}

We first discuss the current status of the four patterns. One sees that, {\bf T1} and {\bf T2} are one-parameter patterns for the lepton mixing matrix, depending only on the parameter $\phi$, to be adjusted to reproduce all angles and the Dirac CP phase $ \delta $. Thus, all the mixing angles and $\delta$ are strongly correlated.
The allowed range of $\phi$ can be obtained by requiring 
all the mixing angles and Dirac phase $\delta$ to lie within their current $3\sigma$ experimental ranges~\cite{deSalas:2017kay}.
Thus, we find $ \phi/\pi \in [0.749, 0.779]$ and $ [0.218, 0.251]$ for {\bf T1} and {\bf T2}, respectively\footnote{Note that the expression for oscillation parameters in Eq.~\ref{eq:BL1} ({\bf T1}), Eq.~\ref{eq:BL2} ({\bf T2}), Eq.~\ref{eq:bl3} ({\bf T3}), Eq.~\ref{eq:bl4} ({\bf T4}) are approximate, i.e. they are truncated at certain order in $\lambda$. The numerical values quoted and used throughout this work are obtained from the full exact expressions.}.

On the other hand {\bf T3} and {\bf T4} depend on two parameters, $\psi$ and $\phi$.
The allowed parameter ranges for $\psi$ and $\phi$ can be found by fitting the current oscillation data with respect to them as shown in Fig.~\ref{fig:phi-psi}.
This Figure provides updated variants of the Figure given in~\cite{Roy:2014nua}. As seen from Fig.~\ref{fig:phi-psi}, current oscillation data
restrict the range of both $\psi$ and $\phi$ to narrow ranges. In particular the value of $\psi$ is restricted to a small band around $\psi = 3$.
In order to easily compare all the four bi-large cases on the same footing, henceforth we will fix $\psi=3$ for both {\bf T3} and {\bf T4} ansatze.
This allows us to put all the results for the four bi-large ansatze in a common figure.
It should be noted however that $\psi=3$ is merely a convenient benchmark choice. Other values of $\psi$ within the narrow allowed band of Fig.~\ref{fig:phi-psi} are equally allowed and will lead to different predicted values for the mixing parameters. In this work we will only discuss the implications of picking the benchmark value $\psi = 3$.

\begin{figure}[!htbp]
\centering
\begin{center}
%
\includegraphics[height=8cm,width=10cm]{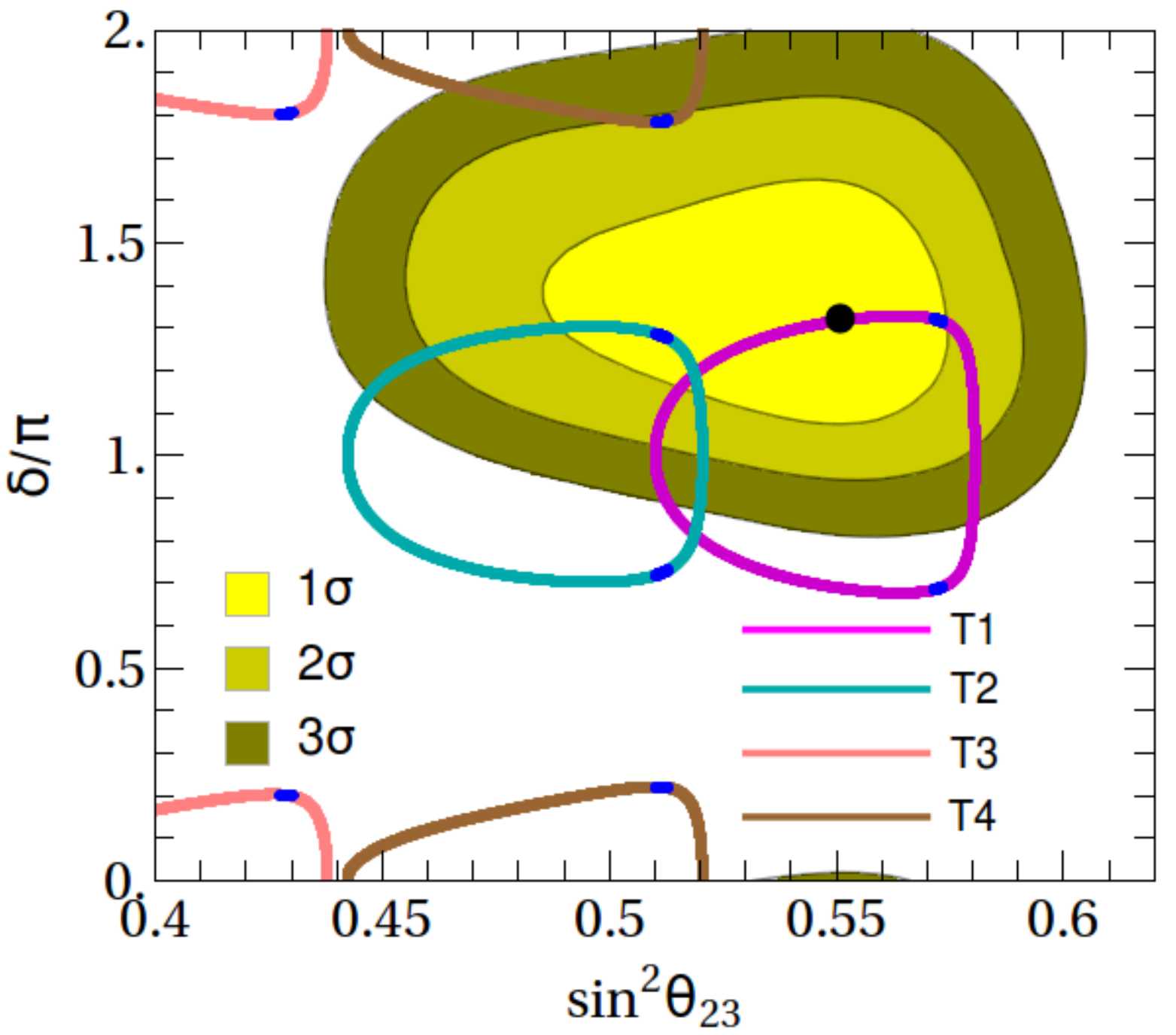}
 \end{center}
\caption{\footnotesize Confronting bi-large mixing schemes with current oscillation data: the filled regions indicate the allowed values in the $\sin^2\theta_{23} - \delta $ plane from the latest global-fit~\cite{deSalas:2017kay}, with the current best-fit point shown by a black dot.
The contours in magenta, cyan, pink, brown correspond to the {\bf T1}, {\bf T2}, {\bf T3}, {\bf T4} scenarios, respectively,
without imposing constraints from the reactor angle $\theta_{13}$.
Imposing the $3\sigma$ bound of the reactor mixing angle leads to the dark-blue patches in the curves.
}
\label{fig:BL12}
\end{figure}

In Fig.~\ref{fig:BL12}, we show the parameter regions allowed by the current global-fit of neutrino
oscillation data in the $\sin^2\theta_{23} - \delta $ plane at $ 1\sigma $ (yellow), $ 2\sigma $ (dark-yellow) and $3\sigma $ (darker-yellow) confidence level. These are indicated by the filled contours, respectively.
The predictions for $\sin^2\theta_{23} $ and $\delta $ are shown
by the magenta, cyan, pink and brown curves for the {\bf T1}, {\bf T2}, {\bf T3} and {\bf T4}, respectively.
However, we emphasize here that the numerical analysis performed for the different bi-large schemes are based on exact formulae and not on the leading order approximations given at the beginning of this section.

Note that each bi-large ansatz predicts two values of $ \delta $, symmetrically placed across the $\delta = \pi$ line.
Since the current oscillation data disfavors $\delta < \pi$ at more than 3$\sigma$, in what follows we only take the upper predicted value of the $\delta $  for the various bi-large scenarios.
From the figures one can see that the predicted values (the one in upper half of CP Plane) of the scenarios {\bf T1}, {\bf T2} and  {\bf T4} all have good agreement with the latest global-fit of neutrino oscillation data. However, the predicted value of {\bf T3} case is more than 3$\sigma$ away from the current value, clearly disfavoured by the latest global-fit~\cite{deSalas:2017kay}. Thus, henceforth we will disregard {\bf T3} as a true value.

Notice that the dark-blue marks in the curves are obtained by requiring $ \sin^2\theta_{13} $ to lie within its current 3$ \sigma $
range \cite{deSalas:2017kay}.
For example, one sees that the solutions of $\sin^2\theta_{23},\delta $ with $(\sin^2\theta_{23},\delta)$ $ \sim(0.57, 1.32\pi) $, $(0.51, 1.28\pi)$ and $\sim (0.51, 1.78\pi) $ are allowed by the current oscillation data for {\bf T1},  {\bf T2} and  {\bf T4}, respectively. Having discussed the current status of the different bi-large mixing schemes, in the next section we examine their testability at the upcoming DUNE experiment.


\section{Testing bi-large Scenarios at DUNE}
\label{sec:test}

\begin{figure}[!htbp]
\includegraphics[height=8cm,width=10cm]{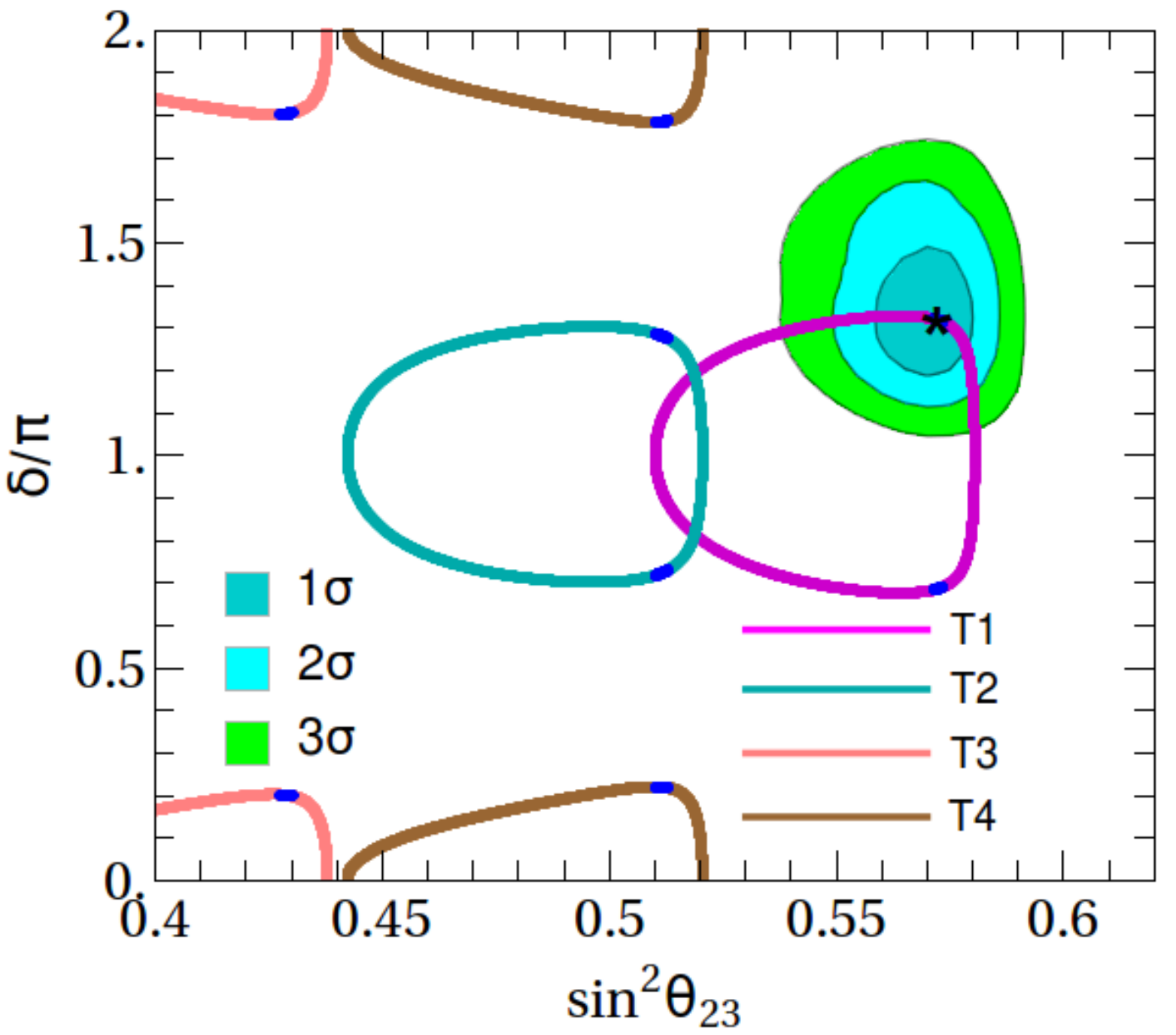}
 \caption{\footnotesize  Capability of DUNE to test one bi-large proposal against others.  Here {\bf T1} prediction is taken as true value, as shown by the `star-mark',  against which the other proposals are tested. The filled-color contours represent the DUNE sensitivity contours after 3.5 + 3.5 years of run.
 }
\label{fig:DUNE-T1}
\end{figure}

Adopting these values  of $(\sin^2\theta_{23},\delta )$ as the seed points (see section \ref{sec:current} for details), we examine sensitivity regions of DUNE. We show the allowed parameter space  considering $ 1\sigma $ (cyan), $ 2\sigma $ (light-cyan) and $3\sigma $ (green) filled-color contours.
From Fig.~\ref{fig:DUNE-T1}, one can see that if  {\bf T1} predicted value is true value, then DUNE after 3.5 + 3.5 years running time can rule out all the other bi-large schemes at more than 3$\sigma$.


\begin{figure}[!htbp]
\centering
\includegraphics[height=8cm,width=10cm]{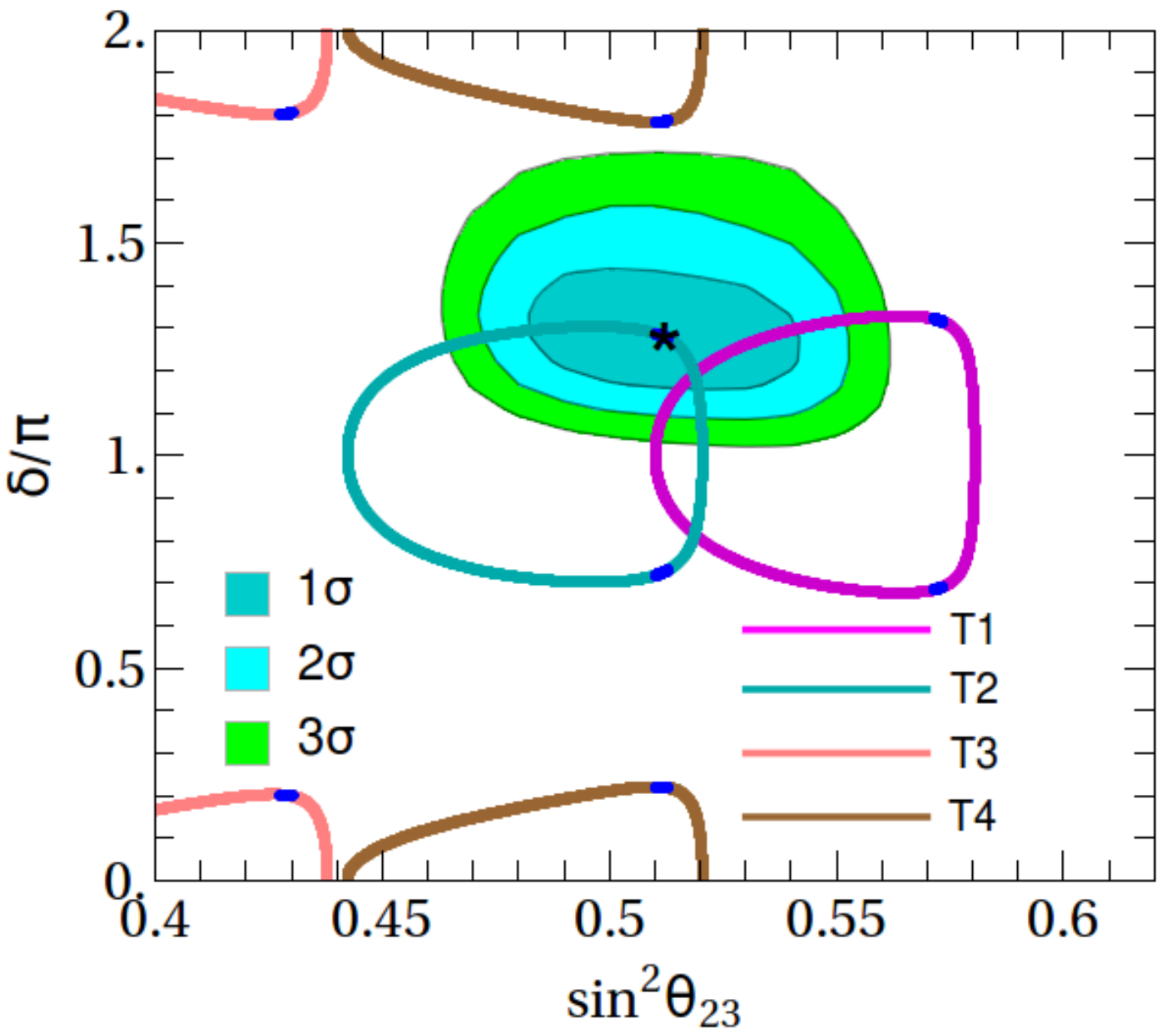}
\caption{\footnotesize  Capability of DUNE to test one bi-large proposal against others. Here {\bf T2} prediction is taken as true value, as shown by the `star-mark', against which the other proposals are tested. The filled-color contours represent the DUNE sensitivity contours after 3.5 + 3.5 years of run.
  }
\label{fig:DUNE-test-T2}
\end{figure}

From Fig.~\ref{fig:DUNE-test-T2}, one can see that if the {\bf T2} predicted value is the true value, then DUNE after 3.5 + 3.5 years run can rule out all of the other bi-large schemes at more than 3$\sigma$. 
The {\bf T1} and {\bf T3} cases would be ruled out as their predicted value for $ \sin^2\theta_{23} $
are very different, while for the case of {\bf T4} the CP phase is very different. As can be seen from Fig.~\ref{fig:DUNE-test-T2} DUNE has enough sensitivity to the CP phase to rule out {\bf T4} as well.

\begin{figure}[!htbp]
\centering
\includegraphics[height=8cm,width=10cm]{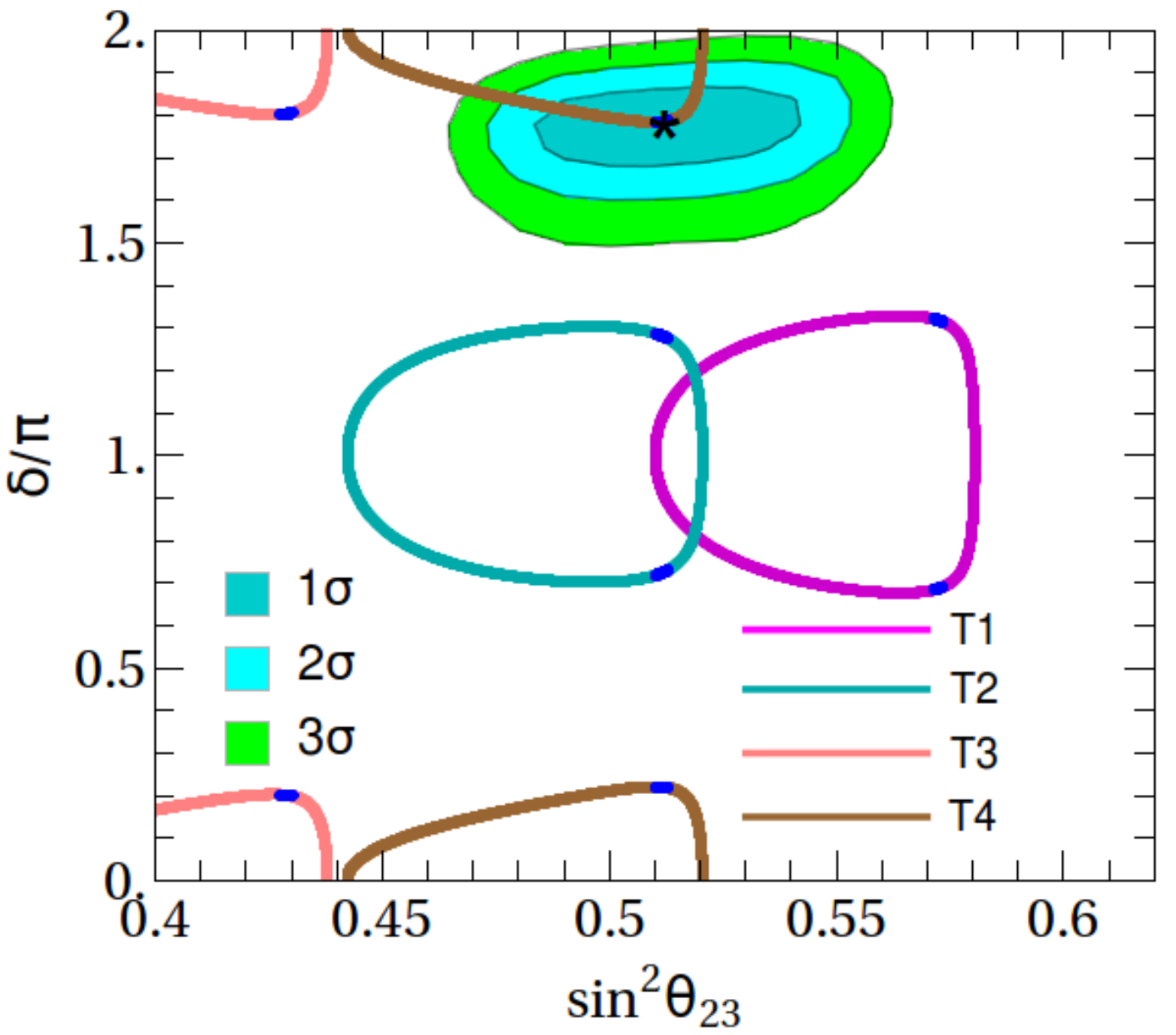}
         \label{fig:bp7}
 \caption{\footnotesize  Capability of DUNE to test one bi-large proposal against others. Here {\bf T4} prediction is taken as true value, as shown by the `star-mark', against which the other proposals are tested. The filled-color contours represent the DUNE sensitivity contours after 3.5 + 3.5 years of run.
 }
\label{fig:DUNE-test-T4}
\end{figure}

By adopting the {\bf T4} predicted value of $\sin^2\theta_{23},\delta $  as our true value, we present DUNE's sensitivity region in Fig.~\ref{fig:DUNE-test-T4}.
We find that, again, in this case DUNE has sensitivity to rule out all the other schemes. {\bf T3} would be ruled out by its predicted value for $ \sin^2\theta_{23} $ while for the cases of {\bf T1} and {\bf T2} the CP phase also plays an important role.
For example, in the case of  {\bf T2} the phase gives the main sensitivity.

\begin{figure}[!htbp]
\centering
\includegraphics[height=8cm,width=9.8cm]{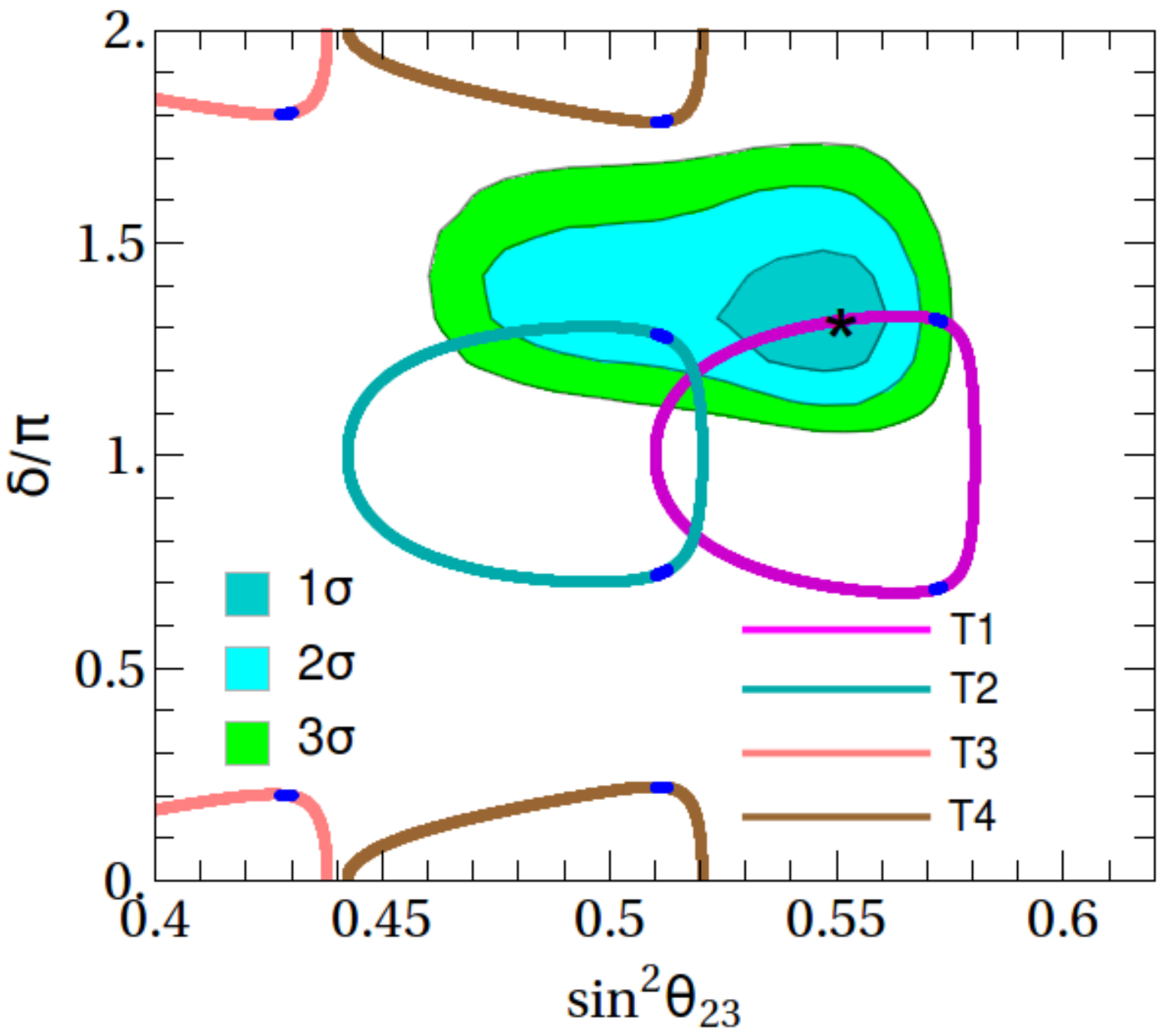}
 \caption{\footnotesize  Capability of DUNE to test bi-large proposals is shown by the filled-color contours. Here current global best-fit value~\cite{deSalas:2017kay} is taken as true value, as shown by the `star-mark', against which the other proposals are tested.
}
\label{fig:DUNE-test-gf}
\end{figure}

Finally, by assuming the current best fit value of the global neutrino oscillation fit as the true value, we show the sensitivity in the $(\sin^2\theta_{23},\delta )$
plane after 3.5 + 3.5 years run  DUNE in Fig.~\ref{fig:DUNE-test-gf}.
One sees that DUNE can rule out the {\bf T3},  {\bf T4} bi-large scheme at more than 3$\sigma$, while the {\bf T1} predicted seed point can be disfavored at more than 2$\sigma$. 
The {\bf T2} scheme can only be disfavoured at the 1$\sigma$ level.

\section{Summary}
\label{sec:summary}

We described four bi-large patterns for the lepton mixing matrix. While {\bf T1} and {\bf T2} are one-parameter patterns, with a single parameter $\phi$ to account for all leptonic mixing angles and Dirac CP phase $\delta$, {\bf T3} and {\bf T4} depend on two parameters $\psi$ and $\phi$.
We have described the current status of the four bi-large patterns obtained by confronting them with current oscillation data.
One of the noticeable points is that when we combine the latest constraint of $\sin^2\theta_{13}$ with the bi-large scheme predictions we obtain very stringent restrictions.
For definiteness in the {\bf T3} and {\bf T4} cases we fixed $\psi=3$, close to the actual allowed value preferred by the analysis of the existing oscillation data. This allows us to put all the results for the four bi-large ansatze on the same footing for comparison. Finally, we described the potential of the DUNE experiment in testing these patterns for the lepton mixing matrix.

\acknowledgments

GJD acknowledges the support of the National Natural Science Foundation of China under Grant Nos 11835013 and 11522546.
The research work of NN was supported in part by the National Natural Science Foundation of China under grant No. 11775231.
 RS and JV are supported by the Spanish grants SEV-2014-0398 and FPA2017-85216-P (AEI/FEDER, UE), PROMETEO/2018/165 (Generalitat Valenciana) and the Spanish Red Consolider MultiDark FPA2017-90566-REDC.

\bibliographystyle{utphys}
\bibliography{bibliography}
\end{document}